\documentclass[twocolumn,showpacs,preprintnumbers,amsmath,amssymb]{revtex4}

\usepackage{graphicx}% Include figure files
\usepackage{dcolumn}% Align table columns on decimal point
\usepackage{bm}% bold math
\usepackage{epsf}
\usepackage{amsmath}

\newcommand{\la}{\langle}
\newcommand{\ra}{\rangle}
\newcommand{\be}{\begin{equation}}
\newcommand{\ee}{\end{equation}}
\newcommand{\bea}{\begin{eqnarray}}
\newcommand{\eea}{\end{eqnarray}}

\begin{document}

\title{Optimal nonequilibrium entanglement of nanomechanical oscillators}

\author{Fernando  Galve and Eric Lutz \\
{\it \small Department of Physics, University of Augsburg, D-86135 Augsburg, Germany}}

\date{\today}

\begin{abstract}
We investigate nonequilibrium entanglement generation in a chain 
of harmonic oscillators with time--dependent linear coupling.  We use optimal control theory to determine the coupling modulation that leads to maximum logarithmic negativity for a pair of opposite oscillators  and show that it corresponds to a synchronization of the eigenmodes of the chain. We further analytically relate the maximum attainable entanglement to the irreversible work done to produce it, thus bridging nonequilibrium entanglement production and nonequilibrium thermodynamics. 
\end{abstract}

\maketitle

Entanglement is a puzzling feature of quantum theory. Entangled states are routinely created and manipulated in dozens of laboratories around the world: Controlled multiparticle  entanglement has for example been achieved for up to eight qubits  in ion traps \cite{lei05,haf05}. In addition, entangled pairs of photons have been prepared over large distances of several kilometers in optical fibers \cite{tit98,mar03} as well as   in free space \cite{asp03,urs07}. Due to their unique properties, entangled states  have lead to many  applications in quantum communication and quantum information theory that are not possible at the classical level,  such as quantum cryptography \cite{eke91} and teleportation \cite{bra93}. And yet, the true nature of entanglement remains elusive, especially in  many--body systems \cite{ami07}.

For quantum many--body systems at thermal equilibrium, macroscopic entanglement  has  recently been related to thermodynamic quantities like the magnetic susceptibility \cite{wie05},   the internal energy \cite{dow04,tot05}, and the heat capacity \cite{gos03}. Some of these theoretical considerations have been successfully confirmed in  experiments with  quantum spin systems \cite{gos03,ver06,rap07,sou08}. In the above examples, equilibrium thermodynamic functions can  thus be regarded as entanglement witnesses  that can be used to detect and characterize entangled states. The situation, however,  is much different away from equilibrium and   the question of how to relate entanglement production to  irreversible thermodynamics  is largely open.

Our purpose in this paper is to determine the maximum attainable entanglement  generated in a nonequilibrium  transformation and  express the optimal value  in terms of the irreversible work done during the process. The concept of irreversible work  plays a fundamental role in the study of nonequilibrium thermodynamics of small systems \cite{bus05}. To be specific, we shall consider
a chain of quantum harmonic oscillators with time--dependent linear coupling. This system  is of direct relevance to the investigation of arrays of nanomechanical resonators \cite{rou01} and ion crystals in multitrap arrangements \cite{wak92}. A notable feature of the linear chain is that its dynamics is exactly solvable. Furthermore, since Gaussian states retain their Gaussian shape, we can rely on exact entanglement measures such as the logarithmic negativity \cite{vid02}.

The equilibrium entanglement properties of a closed harmonic chain with constant coupling have been studied in detail in Ref.~\cite{aud02}. A remarkable result is that individual oscillators  are   only entangled with their immediate neighbors. Contrary to classical correlations, entanglement  hence does not occur over large distances within the chain. Clearly, this property  also persists for adiabatic, that is, infinitely slow transformations of the Hamiltonian, since the chain  is then always in  equilibrium. Recently, Eisert and coworkers have shown that a nonadiabatic transformation -- a sudden switch of the coupling  strength -- can significantly entangle even very distant oscillators, without having to control each of them individually \cite{eis04}. They have  considered the  feasibility of such a scheme in the context of an array of double--clamped coupled nanomechanical oscillators  and discussed the experimental detection of the  entanglement produced.

In the following, we investigate nonequilibrium entanglement generation in the harmonic chain from the point of view of irreversible thermodynamics. We use optimal control theory \cite{kir70} to determine the modulation of the coupling coefficient that leads to maximum entanglement between distant pairs of oscillators. We find that optimal entanglement corresponds to a synchronization of the eigenfrequencies of the linear chain. Moreover, we show that the optimal value can greatly surpass that obtained from a sudden switch of the coupling strength. By further using recent findings on the nonequilibrium work distribution of a single oscillator \cite{def08}, we are able to relate the optimal entanglement to the irreversible work performed during the nonadiabatic transformation.

{\it Harmonic chain.} We consider a closed chain of $N$ harmonic oscillators with identical frequency $\omega_0$ and linear nearest--neighbour coupling with variable intensity $c(t)$:
\begin{equation} \label{eq1}
{H}=\frac{1}{2}\sum_{n=1}^N \Big({p}_n^2+\omega_0^2{q}_n^2+c(t)({q}_{n+1}-{q}_n)^2 \Big) \ .
\end{equation}
%(we will use $m=\hbar=1$ throughtout the paper).
To analyze the Hamiltonian  \eqref{eq1}, it is convenient to introduce the normal mode coordinates $Q_s$ and $P_s$ via,
\begin{equation}
\label{eq2}
{q}_n=\sum_{s=1}^N e^{2\pi i sn/N} \frac{Q_s}{\sqrt{N}}, \hspace{.6cm} {p}_n=\sum_{s=1}^Ne^{2\pi i sn/N}\frac{P_s}{\sqrt{N}} \ .
\end{equation}
 The Hamiltonian of the chain can then  be rewritten as a sum of independent oscillators with time--modulated frequencies, $\omega^2_s(t)=\omega_0^2+4 c(t)\sin^2(\pi s/N)$, 
\begin{equation} \label{eq3}
{H}=\frac{1}{2}\sum_{s=1}^N \left({P}_s{P}_s^\dagger+\omega_s^2(t){Q}_r {Q}_s^\dagger\right) \ ,
\end{equation} 
where we have used the property, $Q_{-s}=Q_s^\dagger$, $P_{-s}=P_s^\dagger$. The linear Heisenberg  equations of motion for $P_s(t)$ and $Q_s(t)$  can readily be written down and solved \cite{eis04}.

The Gaussian states of the linear chain are fully characterized by the symmetric $2 N \times 2N$ covariance matrix $\Gamma$, whose elements are defined by $\Gamma_{q_nq_m} = 2 \mbox{Re} \la q_nq_m \ra$, $\Gamma_{q_np_m} = 2 \mbox{Re} \la q_np_m \ra$ and $\Gamma_{p_np_m} = 2 \mbox{Re} \la p_np_m \ra$. We have  omitted the first moments, since they do not influence the entanglement properties.
Assuming for the time being that the chain is initially in the ground state, and that therefore $\Gamma_{q_nq_m}(0)\, \omega_0= \Gamma_{p_np_m}(0)/\omega_0 = \delta_{n,m}$, $\Gamma_{q_np_m}(0) =0$,   the matrix elements can be written in terms of the second moments of the normal modes as,
\be
\label{eq4}
\la{q}_n{q}_m\ra=\frac{1}{N}\sum_{s=1}^Ne^{2\pi i s(n-m)/N}\la{Q}_s{Q}_s^\dagger\ra \ ,
\ee
using $\la{Q}_r{Q}_s^\dagger\ra=\la{Q}_s{Q}_s^\dagger\ra\delta_{r,s}$. Similar expressions are obtained for $\la{q}_n{p}_m\ra$ and $\la{p}_n{p}_m\ra$. 
The covariance matrix is hence entirely determined by the second moments of the normal coordinates. It is known that a time--dependent  oscillator is squeezed when its frequency is changed nonadiabatically, while an adiabatic transformation leads to  no squeezing \cite{jan87}. We can therefore  express the quadratures of the normal oscillators in terms of the squeezing parameter $r_s$ and the rotation angle $\theta_s$ in the form, 
\begin{subequations}
\label{eq5}
\be
\la{Q}_s{Q}_s^\dagger\ra=\frac{1}{2\omega_s}(e^{-2r_s}\cos^2\theta_s+e^{2r_s}\sin^2\theta_s)\ ,
\ee
\be
\la{P}_s{P}_s^\dagger\ra=\frac{\omega_s}{2}(e^{2r_s}\text{cos}^2\theta_s+e^{-2r_s}\text{sin}^2\theta_s)\ ,
\ee
\be
\la{Q}_s{P}_s^\dagger\ra=\text{sinh}(2r_s)\text{sin}\theta_s\text{cos}\theta_s \ .
\ee
\end{subequations}
The time dependence of $\omega_s$, $r_s$ and $\theta_s$ is controlled by the linear coupling coefficient $c(t)$. Standard thermodynamics considers infinitesimal changes that take place infinitely slowly. In the spirit of finite--time thermodynamics \cite{pet99}, we will in the sequel consider infinitesimal changes that can occur at arbitrary rate. We will thus assume that the coupling intensity is small, $c(t) \ll \omega_0^2$, but otherwise arbitrary. Moreover, without loss of generality, we will set the frequency  $\omega_0=1$.

{\it Maximum entanglement.} The  entanglement between any two oscillators $n$ and $m$ in the chain can be characterized with the help of the logarithmic negativity \cite{vid02}, 
\begin{equation}
\label{eq6}
E_N=-\frac{1}{2}\sum_{i=1}^4\log_2\left[\mbox{Min}(1,|l_i|)\right] \ .
\end{equation}
Here,  $\l_i$ are the symplectic eigenvalues of the covariance matrix of the partially transposed reduced density operator of the two oscillators. In the present investigation, we choose two opposite oscillators, $m=N/2+n$. The latter  configuration corresponds to the largest possible distance in the chain. It can also be shown to correspond to the largest value of the logarithmic negativity due to symmetry reasons. It is worth noting that in this case the exponential factor in Eq.~\eqref{eq4} simplifies to $e^{2\pi is(n-m)/N}=(-1)^s$, thus distinguishing even and odd normal modes. Furthermore, in view of the translational invariance along the chain, the logarithmic negativity does not depend on the particular value of $n$. 

The symplectic eigenvalues of the $4 \times 4$ reduced covariance matrix $\Gamma_{m,n}$ can be computed explicitly using the parametrization \eqref{eq5}. A careful examination of their structure reveals that for moderate squeezing the  logarithmic negativity \eqref{eq6} is maximum when the angles $\theta_s$ take special values. These optimal angles are given by  $\theta_s= \pi /4 \,(\mbox{mod} \pi)$ for $s$ even and
$\theta_s= 3\pi /4 \, (\mbox{mod} \pi)$ for $s$ odd. Hence, for fixed squeezing, maximum entanglement is achieved when both even and odd oscillators are synchronized.  From Eqs.~\eqref{eq5}, it follows that, at the optimal angles, $\la{Q}_s{Q}_s^\dagger\ra=\cosh(2r_s)/2\omega_s$, $\la {P}_s{P}_s^\dagger\ra=\cosh(2r_s)\, \omega_s/2$ and $\la{Q}_s{P}_s^\dagger\ra=(-1)^s\sinh(2r_s)/2$. Combining Eqs.~\eqref{eq4} and \eqref{eq5},  with $\omega_s \simeq \omega_0$, we then find that $\la q_n^2\ra = \la p_n^2 \ra =a $ and $\la q_nq_m\ra = \la p_n p_m\ra=b$. As a result,  the symplectic eigenvalues in Eq.~\eqref{eq6} are simply given by,
\begin{equation}
\label{eq7}
|l_{_2^1}|=\sqrt{[a\mp b-(c\mp d)][a\pm b+c\pm d]}\ ,
\end{equation}
with $c= \la q_np_n\ra$ and  $d= \la q_np_m\ra$. We therefore obtain that the maximum attainable logarithmic negativity is:

\begin{equation}
\label{eq8}
E_N^{max}=\!-\frac{1}{2}\text{log}_2\!\left[\!\left(\frac{2}{N}\!\!\sum_{s,odd}e^{-2r_s}\right)\!\!\left(\frac{2}{N}\!\!\sum_{s,even}e^{-2r_s}\right)\!\right]\!.\!
\end{equation}
Equation \eqref{eq8} shows that  entanglement is zero for vanishing squeezing and increases with increasing squeezing. By comparing the simplified symplectic eigenvalues \eqref{eq7} with their general expression, we further  find that the optimal angles are valid provided the condition $c\sum_s \exp(2r_s) /4 \ll N$ is satisfied. As we shall discuss below, this condition is not very restrictive. 

\begin{figure}
\includegraphics[width=8.5cm]{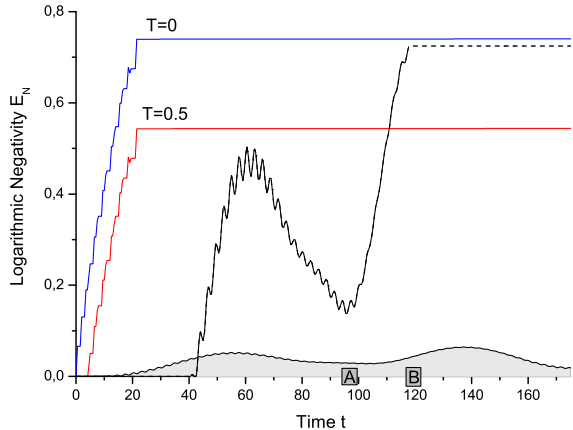}
\caption{Logarithmic negativity, Eq.~\eqref{eq6}, for opposite oscillators  as a function of time for different modulations of the coupling strength $c(t)$ between $c=0$ and $c=0.05$.  The lower shaded curve corresponds to a sudden switch at $t=0$, while the curve in the center is generated by the optimal modulation shown in Fig.~\ref{fig3}. The two upper curves are the maximum attainable entanglement for $T=0$ and $T=0.5$.}\label{fig1}
\end{figure}

\begin{figure}
\includegraphics[width=8.7cm]{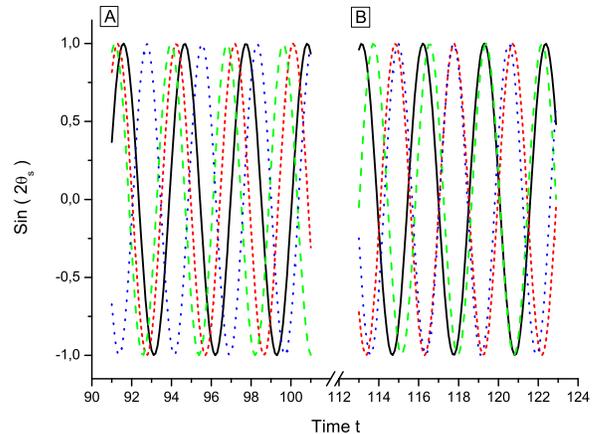}
\caption{Angles $\theta_s$  of the normal modes as a function of time for the two intervals A and B shown in Fig.~\ref{fig1} (due to the symmetry of the chain, angles are pairwise equal). The synchronization of  even and odd modes is clearly visible  at the center of interval B and completely absent in interval A.}\label{fig2}
\end{figure}

\begin{figure}
\includegraphics[width=8.cm]{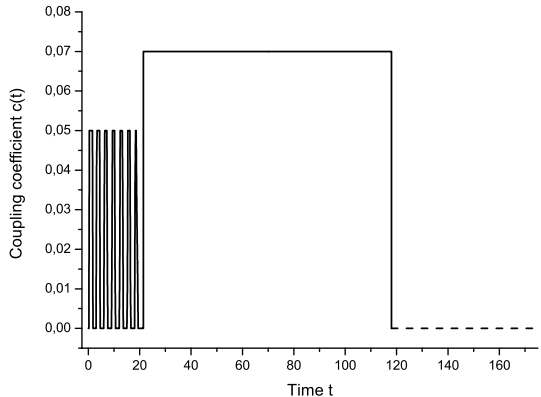}
\caption{Optimal coupling modulation $c(t)$ that generates the logarithmic negativity shown in the center of Fig.~\ref{fig1}.}\label{fig3}
\end{figure}

{\it Optimal control theory.} Optimal control theory (OCT) is a powerful method that allows to determine the function that optimizes a given cost functional  \cite{kir70}. It can be regarded as a generalization of the calculus of variations and has a long tradition in finite--time thermodynamics \cite{pet99}. We apply Pontryagin's principle to find the function $c(t)$ that optimizes the entanglement between opposite oscillators. To this end, we take as the cost functional to minimize, the argument of the maximum attainable logarithmic negativity \eqref{eq8}, $\sum_{s,odd}\exp(-2r_s)  \sum_{s,even}\exp(-2r_s)$. It is clear that the minimum of this quantity will lead to  maximum entanglement. The constraints given by the equations of motion of the normal modes \eqref{eq2} enter the optimization problem in the form of Lagrange multipliers. The optimal function $c(t)$ is eventually obtained by steepest descent. The results of the numerical implementation of optimal control theory for  a chain of $N=8$ oscillators are summarized in Figs.~\ref{fig1}, \ref{fig2} and \ref{fig3}.

Figure \ref{fig1} shows the logarithmic negativity $E_N$ of oscillators 1 and 5 as a function of time for various coupling intensities $c(t)$. The shaded curve at the bottom corresponds to a sudden switch  from $c= 0$ to $c= 0.05$. The oscillatory curve in the center depicts the entanglement generated using OCT for the same parameters.  The upper curve, on the other hand, is the maximum attainable entanglement at zero temperature, Eq.~\eqref{eq8}, when the angles $\theta_s$ of the normal oscillators are perfectly synchronized at all times. We observe that the entanglement produced by the sudden switch is one order of magnitude  {\it smaller} than the maximum attainable value in the present situation. Moreover, we note that the optimal logarithmic negativity (almost) reaches the maximum attainable value $E_N^{max}$. 
Figure \ref{fig2} shows the angles of the normal oscillators, corresponding to the optimal modulation, in two short time intervals: the first  around the peak of highest entanglement (denoted by B in Fig.~\ref{fig1}) and  the second around a minimum of the logarithmic negativity (interval A).  The (nearly perfect) synchronization of the even and odd modes at the center of the interval is clearly visible in case B, while it is  absent in case A. 
 The optimal modulation $c(t)$  is plotted in Fig.~\ref{fig3} and consists of  three different parts. The first part is a sequence of square pulses between $c= 0$ and $c= 0.05$. The latter  is the  result of the optimization procedure that tries to maximally squeeze each normal oscillator in order to minimize the argument of  Eq.~\eqref{eq8}; the maximal entanglement $E_N^{max}$ is determined by the duration of this sequence. In the second part, the coupling $c(t)$ is kept constant allowing the angles to synchronize \cite{rem};  the synchronization time is directly controlled by  the value of $c$: a larger coupling will lead to a faster synchronization. Since each  normal mode evolves with a different frequency, the synchronization of the angles cannot persist in time. However, by stopping the modulation of $c(t)$ (dashed line in Fig.~\ref{fig3}), once the biggest  entanglement has been attained, the peak value  can be maintained  (see  dashed line in Fig.~\ref{fig1}).
 
{\it Dissipative work.} We next relate the nonequilibrium entanglement produced by a nonadiabatic switching of the coupling to the average dissipated work. The dissipated work is a central concept in thermodynamics and is intimately  connected to the irreversible entropy production \cite{bus05}; it is defined as the difference between the total work and the reversible work (or  free energy difference), $W_{dis}= W- \Delta F$. For an adiabatic  transformation, $W_{dis}=0$ and the total work is given by $ \Delta F$. The dissipated work for a frequency--modulated normal oscillator of the chain can be simply expressed in terms of the frequency  and the squeezing parameter at time $t$ \cite{gal08}:
\begin{equation}
\label{eq9}
W_{dis,s}=\omega_s\text{sinh}^2r_s \ .
\end{equation}

\noindent The above equation indicates that there is a one--to--one correspondence between the dissipated work  $W_{dis,s}$ and  the squeezing parameter $r_s$: small squeezing yields  small dissipated work, $W_{dis,s}\simeq\omega_s r_s^2$, whereas large squeezing leads to an exponential increase of the dissipated work,   $W_{dis,s}\simeq\omega_s \exp(2r_s)/4$. The total work dissipated in the chain is  given by the sum over all eigenmodes, $W_{dis} = \sum_s W_{dis,s}$. By combining Eqs.~\eqref{eq8} and \eqref{eq9}, we can now directly relate the maximum attainable logarithmic negativity to the irreversible work required to generate it. This relationship is of importance for several reasons: First, from a fundamental point of view, it bridges nonequilibrium entanglement production with nonequilibrium thermodynamics; second,  it gives us the maximum amount of entanglement that can be produced by dissipating a given amount of work, in the ideal situation where all the phases are synchronized. One could for instance dissipate considerably more work by highly squeezing  few oscillators in the chain, irrespective of their angles, and produce entanglement far below the maximum possible value. In that respect, this relationship is similar in spirit to the Carnot efficiency of a heat engine that gives the maximum efficiency that can be achieved   for given temperatures, in an ideal cycle.

{\it Effect of temperature.} Our discussion so far was limited to a chain at zero temperature. Finite temperatures can easily be  taken into account by multiplying the initial covariance matrix $\Gamma(0)$ by the factor $f(T)=\coth(\omega_0/2T)$. The new symplectic eigenvalues are then  merely  multiplied by the factor $f(T)$. As a consequence, the maximum attainable logarithmic negativity is still given by its expression \eqref{eq8}, with the argument now multiplied by $f(T)$. Since $f(T)\geq1$, temperature reduces the maximal value of entanglement, as expected. Figure \ref{fig1} shows the maximal logarithmic negativity for $T=0.5$. 
A good estimate of the highest temperature $T_m$ at which entanglement vanishes can be obtained by replacing the squeezing parameters of the different eigenmodes by the average squeezing, $R= \sum_s r_s /N$. We then find that entanglement survives up to $T_m=\omega_0/\ln(\coth(2R))$.

{\it Discussion.}
One of the main  limitations that have to be considered in real systems is the validity of the harmonic approximation. As the oscillators are squeezed, their energy increases and anharmonicities may appear. For the optimal modulation of Fig.~\ref{fig3}, with a squeezing sequence up to $t=20$, $R= 0.4$ and $W_{dis}=1.2$ (in units of $\hbar$). This leads to an increase of 30\% of the total energy of the chain from $E_0=4$ to $E_0+W_{dis}=5.2$. A tenfold energy increase  is achieved with a squeezing sequence up to $t=85$, yielding $R=1$ and $E_N^{max}=1.6$. These parameters are well within the bounds of the criterion given below Eq.~\eqref{eq8}. A second limitation is environment induced decoherence. Fortunately, as shown in Ref.~\cite{eis04}, entanglement in the chain is particularly robust against external noise sources at low temperatures.

This work was supported by  the
Emmy Noether Program of the DFG (Contract LU1382/1-1) and the
cluster of excellence Nanosystems Initiative Munich (NIM). EL would like to acknowledge the hospitality of the Czech Technical University in Prague.


\begin{thebibliography}{99}
\bibitem{lei05} D. Leibfried {\it et al.}, Nature {\bf 438}, 639 (2005).
\bibitem{haf05} H. H\"affner {\it et al.}, Nature {\bf 438}, 643 (2005). 
\bibitem{tit98} W. Tittel {\it et al.}, Phys. Rev. Lett. {\bf 81}, 3563 (1998).
\bibitem{mar03} I. Marcikic {\it et al.}, Nature {\bf 421}, 509 (2003).
\bibitem{asp03} N. Aspelmeyer {\it et al.}, Science {\bf 301}, 621 (2003).
\bibitem{urs07} R. Ursin {\it et al.}, Nature Physics {\bf 3}, 481 (2007).
%\bibitem{vil08} P. Villoresi {\it et al.}, New J. Phys. {\bf 10}, 033038 (2008).
\bibitem{eke91} A. Ekert, Phys. Rev. Lett. {\bf 67}, 661 (1991).
\bibitem{bra93} C.H. Bennett {\it et al.}, Phys. Rev. Lett. {\bf 70}, 1895 (1993).
\bibitem{ami07} L. Amico, R. Fazio, A. Osterloh, and V. Vedral, Rev. Mod. Phys.  {\bf 80}, 517 (2008). 
\bibitem{wie05} M. Wie\'sniak, V. Vedral. and C. Brukner,   New J. Phys. {\bf 7}, 258 (2005).
\bibitem{dow04} M.R. Dowling, A.C. Doherty, and S.D. Bartlett, Phys. Rev. A {\bf 70}, 062113 (2004).
\bibitem{tot05}  G. T\'oth, Phys. Rev. A {{\bf 71}, 010301(R) (2005).
%\bibitem{and06} J. Anders, D. Kaszlikowski, C. Lunkes, T. Ohshima, and V. Vedral,  New J. Phys. {\bf 8} 140 (2006). 
\bibitem{gos03} S. Ghosh {\it et al.}, Nature {\bf 425}, 48 (2003).
\bibitem{ver06} T. V\'ertesi and E. Bene, Phys. Rev. B {\bf 73}, 134404 (2006).
\bibitem{rap07} T.G. Rappoport { \it et al.}, Phys. Rev. B {\bf 75}, 054422 (2007).
\bibitem{sou08}  A.M. Souza {\it et al.}, Phys. Rev. B {\bf 77}, 104402 (2008).
\bibitem{bus05} F. Ritort,  S\'eminaire Poincar\'e {\bf 2},  193 (2004).
\bibitem{rou01} M. Roukes, Phys. World {\bf 14}, 25 (2001); H.G. Craighhead, Science {\bf 290}, 1532 (2000).
\bibitem{wak92} G. Ciaramicoli {\it et al.}, Phys. Rev. Lett. {\bf 91}, 017901 (2003).
\bibitem{vid02} G. Vidal and R.F. Werner, Phys. Rev. A {\bf 65}, 032314 (2002).
\bibitem{aud02} K. Audenaert {\it et al.}, Phys. Rev. A {\bf 66}, 042327 (2002).
\bibitem{eis04} J. Eisert {\it et al.}, Phys. Rev. Lett. {\bf 93}, 190402 (2004); M.B. Plenio {\it et al.}, New. J. Phys. {\bf 6}, 36 (2004).
\bibitem{kir70} D.E. Kirk, {\it Optimal Control Theory}, (Prentice--Hall, Englewood Cliffs, 1970).
\bibitem{def08} S. Deffner and E. Lutz, Phys. Rev. E {\bf 77},  021128 (2008).
\bibitem{jan87} J. Janszky and Y.Y. Yushin, Opt. Commun. {\bf 59}, 151 (1986);
R. Graham, J. Mod. Opt. {\bf 34}, 873 (1987). 
%\bibitem{gro91} J. Grochmalicki and M. Lewenstein, Phys. Rep. {\bf 208}, 189 (1991).
\bibitem{pet99} R.S Berry {\it et al.}, {\it Thermodynamic Optimization of Finite-Time Processes}, (Wiley, New York, 1999).
\bibitem{rem} It does not seem to be possible to simultaneously squeeze and synchronize the normal modes of the chain by modulating the sole coupling intensity $c(t)$. 
\bibitem{gal08} F. Galve and E. Lutz (to be published). 
}
\end{thebibliography}
\end{document}